# On a logical model of combinatorial problems

A.D. Plotnikov

## Abstract

The paper proposes a logical model of combinatorial problems, also it gives an example of a problem of the class NP that can not be solved in polynomial time on the dimension of the problem.

## 1. Statement of the problem

Suppose we have a *n*-set $A = \{a_1, a_2, ..., a_n\}$. The problem of constructing a sample $S \subseteq A$ of elements of *A*, satisfies a specified conditions, is called combinatorial. Elements of the set *A* can be numbers, symbols, geometric objects, etc.

Logic is the natural language of mathematics. Therefore, the construction of logical models of combinatorial problems helps to better understand the features of a problem, estimate the possible ways and the complexity of the solving.

Any mass problem is characterized by some list of parameters (in our case, this is the set *A*) and a predicate *P(S)*, which determines the properties of the solution *S* is required to satisfy [2, 3]. In complexity theory introduces the concept of problems, which form the class NP.

The problem belongs to the class NP if the solution can be checked for the time described by a polynomial $p = f(n)$ on the problem dimension *n*. There are many problems that belong to the class of NP [1, 2], which can be solved in the time, is also described by a polynomial $t = \varphi(n)$. The set of all these problems form the class P. The central question of complexity theory is the problem of the relation of classes P and NP, i.e, P = NP or P ≠ NP?

The purpose of this paper is to offer a general logical model of combinatorial problems, the solution of which is a disordered sample, as well as to estimate the complexity of solving certain problems.

## 2. The general logical model

Usually, each combinatorial problem is defined as a triple *(A, P(S), W(S))*, where $A = \{a_1, a_2, ..., a_n\}$ is *n*-set of solution elements, *P(S)* is a predicate that determines whether some subset $S \subseteq A$ satisfies conditions of the problem, *W(S)* is a cost function of *S*.

Each subset $S \subseteq A$ we associate with the *n*-dimensional Boolean tuple $B = \{b_1, b_2, ..., b_n\}$, where $b_i = 1$ (*i* = 1, 2,..., *n*) if $a_i \in S$ and $b_i = 0$ otherwise.

The predicate *P(S)* equals to 0 or 1 for each concrete subset S. Therefore, the value of the predicate *P(S)* for each tuple *B* defines the value of a Boolean function *f(B)*, depending on *n* of Boolean variables. Such Boolean function is called a pointer of feasible solutions (PFS).

Thus, the combinatorial problem can be represented as the three (*A, f(B), W(B)*), where *W(B) = W(S)*.

Fig. 1 schematically illustrates the proposed model.

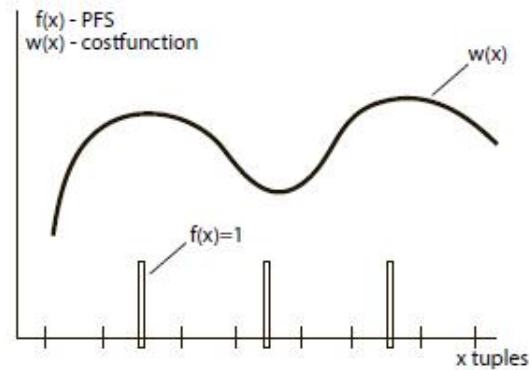

Fig. 1:

Consider a few examples.

**Example 1.** *(The maximum independent set problem).* Let there be an undirected graph $G = (V, E)$, where we want to find the maximum independent set of vertices. Here $V = \{v_1, v_2, ..., v_n\}$ is the set of graph vertices, $E = \{e_1, e_2, ..., e_m\}$ is the set of graph edges.

It is obvious that here $A = V$ is the set of solution elements, the predicate *P(S)* is defined by a procedure that determines whether a subset of vertices $S \subseteq V$ is independent, i.e. whether the vertices of *S* are pairwise non-adjacent. The cost function *W(S)* calculates the number of elements in *S*.

Each subset $S \subseteq V$ we associate with a Boolean tuple $X = \{x_1, x_2, ..., x_n\}$, where $x_i = 1$ if $v_i \in S$ and $x_i = 0$ otherwise. Then the predicate *P(S) = P(X)* defines a Boolean function *f(B)*. We will define *f(B)*.

Each vertex $v_i \in V$ we associate with the conjunction of $C_i = x_i \,\&\, \overline{x_{i_1}} \,\&\, \overline{x_{i_2}} \,\&\, ... \,\&\, \overline{x_{i_k}}$, where $\overline{x_{i_1}}, \overline{x_{i_2}}, ..., \overline{x_{i_k}}$ are variables corresponding to all vertices of $\overline{v_{i_1}}, \overline{v_{i_2}}, ..., \overline{v_{i_k}}$, adjacent to a vertex $v_i$ graph *G*. Obviously, the disjunction

$$D = \bigvee_{i=1}^{n} C_i$$

is the desired pointer of feasible solutions *f(B)*.

Thus, the solution to the problem is to find a tuple $B$, which has the maximum number of unities, on which the function $f(B)$ is equals to one (true).

For example, suppose, there is a graph $G = (V, E)$, shown in Fig. 2, in which we must find the maximum number of independent vertices.

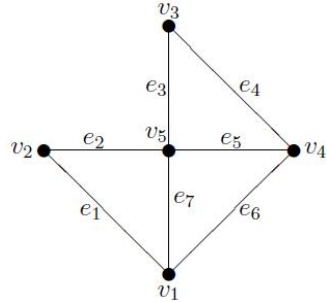

Fig. 2:

Each vertex $v_i \in V$ we associate with a Boolean variable $x_i$ ($i = 1, 2, ..., 5$), where $x_i = 1$ if the vertex $v_i$ belongs to the set of independent vertices and $x_i = 0$ otherwise. We find: $C_1 = x_1 \bar{x}_2 \bar{x}_4 \bar{x}_5$, $C_2 = x_2 \bar{x}_1 \bar{x}_5$, $C_3 = x_3 \bar{x}_4 \bar{x}_5$, $C_4 = x_4 \bar{x}_1 \bar{x}_3 \bar{x}_5$, $C_5 = x_5 \bar{x}_1 \bar{x}_2 \bar{x}_3 \bar{x}_4$.

So, we must find the values of Boolean variables $B = \{x_1, x_2, ..., x_n\}$, corresponding to the vertices $v_1, v_2, ..., v_n$ of the graph $G$, for where the function

$$W(B) = \sum_{i=1}^{n} x_i$$

is maximum provided that the Boolean function
$$f(B) = x_1 \bar{x}_2 \bar{x}_4 \bar{x}_5 \lor x_2 \bar{x}_1 \bar{x}_5 \lor x_3 \bar{x}_4 \bar{x}_5 \lor x_4 \bar{x}_1 \bar{x}_3 \bar{x}_5 \lor x_5 \bar{x}_1 \bar{x}_2 \bar{x}_3 \bar{x}_4 = 1.$$

**Example 2.** *(The Hamiltonian cycle problem).* Let there be an undirected graph $G = (V, E)$, for which it is necessary to find a Hamiltonian cycle if it exists. The graph $G$ is Hamiltonian if there exists a simple cycle that includes all the vertices of the graph.

It is obvious that here $A = E$ is the set of solution elements – the set of graph edges. Predicate $P(S)$ defines procedure that determines whether the basic condition of Hamiltonian graph is executed, namely, a subset of edges $S \subseteq E$ is a subgraph, in which each vertex is incident with at most two edges. The cost function $W(S)$ determines whether a subgraph $S$ is a simple cycle of the graph $G$.

In this case it is convenient initially to construct a function $\bar{f}$, the inverse to the pointer of feasible solutions $f$.

We associate each edge $e_j \in E$ ($j = 1, 2, \ldots, m$) with Boolean variable $x_j$. We believe $x_j = 1$ if the edge $e_j$ belongs the selected set of edges S, and $x_j = 0$ otherwise.

The function $\overline{f}$ is unity on such the tuple of Boolean variables $x_j$, which correspond to the collection of three or more edges incident to the same vertex of G. For example, if the vertex $v_j \in V$ of the graph G is incident $d(v_j) \geq 3$ edges $e_{j1}, e_{j2}, \ldots, e_{jd(v_j)}$ then there is

$$\frac{d(v_j)!}{3!(d(v_j)-3)!}$$

conjunctions of the form

$$C_j = x_{j1} x_{j2} x_{j3},$$

which correspond to tuples containing three edges that incident to the same vertex. Obviously, one should not write conjunction with more than three variables, as they are absorbed by conjunctions of the three variables.

Writing down all such conjunctions for vertices with local degree equals to or more than three, we obtain the inverse function for PFS. For the graph shown in Fig. 2, we have:

| Vertex | Conjunctions |
|---|---|
| $v_1$ | $x_1 x_6 x_7$ |
| $v_4$ | $x_4 x_5 x_6$ |
| $v_5$ | $x_2 x_3 x_5 \vee x_2 x_3 x_7 \vee x_2 x_5 x_7 \vee x_3 x_5 x_7$ |

Therefore, the function $\overline{f}(B)$, inverse for PFS, would be:

$$\overline{f}(B) = x_1 x_6 x_7 \vee x_4 x_5 x_6 \vee x_2 x_3 x_5 \vee x_2 x_3 x_7 \vee x_2 x_5 x_7 \vee x_3 x_5 x_7.$$

Hence, we have the final form for the pointer of feasible solutions:

$$f(B) = (\overline{x}_1 \vee \overline{x}_6 \vee \overline{x}_7)(\overline{x}_4 \vee \overline{x}_5 \vee \overline{x}_6)(\overline{x}_2 \vee \overline{x}_3 \vee \overline{x}_5)(\overline{x}_2 \vee \overline{x}_3 \vee \overline{x}_7) \& $$
$$\&(\overline{x}_2 \vee \overline{x}_5 \vee \overline{x}_7)(\overline{x}_3 \vee \overline{x}_5 \vee \overline{x}_7)$$

As for the function W(B), then it is given by the procedure which establishes that the considered subgraph is a Hamiltonian cycle. This can be executed by depth-first search (DFS) in the graph.

**Example 3.** *(The satisfiability problem).* In the satisfiability problem, some Boolean function $F(x_1, x_2, \ldots, x_n)$ is given, and requires to establish the existence of such Boolean variables $x_1, x_2, \ldots, x_n$, which deliver a unit value function F. It is generally believed that function F is defined as a conjunctive normal form (CNF).

Obviously, in this case literals of Boolean variables are the elements of the solution, that is, the variables $x_1, x_2, ..., x_n$ with negation or without negation. Literals $x$ and $\overline{x}$ called alternative (contraries). Any tuple of non-alternative literals would represent a feasible solution to the problem. Since each tuple $B$ of $n$ Boolean variables determines the feasible set of literals, then the pointer of feasible solutions $f(B) = 1$ for all tuples. In other words, in this case, the PFS is a constant unity.

The cost function is defined by the given Boolean function, that is, $W(B) = F(x_1, x_2, ..., x_n)$.

## 3. On the complexity of solutions

In the process of solving the problem, we have two stages:

• input of initial data of the problem;

• the proper stage of solving the problem.

The number of symbols, which is required for recording initial data, is the dimension of the problem. To solve the problem, it is obviously necessary at least once "view" all of its initial data. Therefore, the complexity of solving any problem can not be less than $O(n)$, where $n$ is the dimension of the problem. An example of such a problem can be a problem to find the maximum element of the array $M$.

In principle, the problem of the search of the maximum element of an array (PME) $M$ can be represented in the form of our logical model. For example, the binary address of the cell of the array can be considered as a set of abstract Boolean variables. The pointer of feasible solutions in this case is a constant 1. The cost function is given by the comparison procedure given numbers. Clearly, to solve this problem, it is necessary to consider all elements of the array, since we do not know how its values of these elements are arranged in an array.

Next, we consider the following example.

Let there be a Boolean tuple of length 4 (a tetrad) $X = (x_1, x_2, x_3, x_4)$. Each such tuple we associate with the number $T = w_1 + w_2 + w_3 + w_4$ — a weight of the tetrad $X$. Summands $w_i$ ($i = 1, ..., 4$) is calculated as follows:

• if $x_1 = 0$ then $w_1 := 5$ else $w_1 := 13$;
• if $x_1 = 0$ and $x_2 = 0$ then $w_2 := 7$;
• if $x_1 = 0$ and $x_2 = 1$ then $w_2 := 10$;
• if $x_1 = 1$ and $x_2 = 0$ then $w_2 := 12$;
• if $x_1 = 1$ and $x_2 = 1$ then $w_2 := 4$;

- if $x_3 = 0$ then $w_3 := 3$ else $w_3 := 8$;
- if $x_3 = 0$ and $x_4 = 0$ then $w_4 := 2$;
- if $x_3 = 0$ and $x_4 = 1$ then $w_4 := 15$;
- if $x_3 = 1$ and $x_4 = 0$ then $w_4 := 3$;
- if $x_3 = 1$ and $x_4 = 1$ then $w_4 := 17$.

We assume that the values of the summands $x_i$, depending on the specified conditions, are determined by some random process. We believe that the weight of the tetrad $T$ equals to $T = w_1 + w_2 + w_3 + w_4$.

Suppose also that there is a pointer of feasible solutions $f(x_1, x_2, ..., x_n)$, presented in the table below. Furthermore, in this table are calculated weights of the corresponding tetrads.

| $x_1x_2x_3x_4$ | $w_1$ | $w_2$ | $w_3$ | $w_4$ | $W$ | $f$ |
|---|---|---|---|---|---|---|
| 0000 | 5 | 7 | 3 | 2 | 17 | 1 |
| 0001 | 5 | 7 | 3 | 15 | 30 | 0 |
| 0010 | 5 | 7 | 8 | 3 | 23 | 1 |
| 0011 | 5 | 7 | 8 | 17 | 37 | 0 |
| 0100 | 5 | 10 | 3 | 2 | 20 | 1 |
| 0101 | 5 | 10 | 3 | 15 | 33 | 1 |
| 0110 | 5 | 10 | 8 | 3 | 26 | 1 |
| 0111 | 5 | 10 | 8 | 17 | 40 | 0 |
| 1000 | 13 | 12 | 3 | 2 | 30 | 1 |
| 1001 | 13 | 12 | 3 | 15 | 43 | 0 |
| 1010 | 13 | 12 | 8 | 3 | 36 | 0 |
| 1011 | 13 | 12 | 8 | 17 | 50 | 0 |
| 1100 | 13 | 4 | 3 | 2 | 22 | 1 |
| 1101 | 13 | 4 | 3 | 15 | 35 | 0 |
| 1110 | 13 | 4 | 8 | 3 | 28 | 1 |
| 1111 | 13 | 4 | 8 | 17 | 42 | 0 |

It is easy to see that we have a equilibrium Boolean function, that is, on half of tuples, this Boolean function is equals to 0, and the other half equals to 1. The formula for calculating the given Boolean function (the pointer of feasible solutions) has the form:

$$f(x_1, x_2, x_3, x_4) = (\overline{x_1} \vee x_2 \vee \overline{x_3})\overline{x_4} \vee \overline{x_1}x_2\overline{x_3}.$$

Let it is required to find a tuple X of the maximum weight at which $f(X) = 1$.

In general, the considered problem (we call it as "Heavy tuple (HT)") can be formulated as follows.

Suppose we have $n$ Boolean variables $x_1, x_2, ..., x_n$. For simplicity, we assume that $n = 4k$ ($k \geq 1$). For each tuple $\Sigma = (\sigma_1, \sigma_2, ..., \sigma_n)$ of Boolean

variables $x_1, x_2, ..., x_n$, we define its weight $w(\Sigma)$ as the sum of the weights of its tetrads:

$$w(\Sigma) = T_1 + T_2 + ... + T_k.$$

In addition, let be the given the equilibrium Boolean function $f(x_1, x_2, ..., x_n)$, whose value can be computed in polynomial time by $n$: $p_n(\Sigma)$.

It is required to find a tuple $\Sigma = (\sigma_1, \sigma_2, ..., \sigma_n)$ of the maximum weight $W_{max}(\Sigma)$ such that $f(\Sigma) = 1$.

The same problem in the decision form was formulated in [4]. It was also shown that the problem belongs to the class NP. Difference problem "Heavy tuple" from problems, presented in Examples 1 – 3, is the absence of knowledge of the predicate $P(S)$, that is, we do not know the properties that must be satisfied of every feasible solution.

**Theorem 1.** *The problem "Heavy tuple" can not be solved in polynomial time on the dimension of the problem.*

Obviously, the problem HT is formulated as an analogue of the problem of finding the maximum element of the given array (PME).

In fact, in the problem "Heavy tuple", addresses (binary tuples) of the array of feasible solutions are determined by the pointer of feasible solutions f. Calculation of the address in this case can be performed by brute force only, because otherwise we have a problem of the function $f^{-1}$, the inverse of the PFS, that is, find a tuple of values of Boolean variables, where the pointer of feasible solutions equal to 1. The procedure for calculating such an address is not associated with the weight of the array element. Only after obtaining the address of the array element, we find (calculate) its value. The same sequence of operations is used for PME.

Assuming that we can calculate the tuple $\Sigma$, where there is the maximum weight. However such tuple may not be the feasible solution, as the pointer $f(\Sigma) = 0$. Since we do not know the predicate $P(S)$ explicitly, this does not allow to consider other ways to solve than the "preview" of all elements of a given array of feasible solutions, the number of which is not less than $2^{n-1}$. " Q.E.D.

**Corollary 2.** *The class P does not coincide with the class NP, that is, $P \supset NP$.*

## 4. Conclusion

The proposed combinatorial model allows with single point of view to consider not only the well-known combinatorial problems, but also to consider new such as problem "Heavy tuple". The most important consequence of the proposed model is to establish the inequality of classes P and NP on the ex-ample of the problem "Heavy tuple". The conclusions, obtained in the study of this problem, can not be extended to problems, considered in Examples $1-3$, as predicate *P(S)* explicitly specified in them.